# Long Wave Dynamics Along A Convex Bottom

**Ira Didenkulova[1,2], Efim Pelinovsky[2], and Tarmo Soomere[1]**

1) Institute of Cybernetics, Tallinn University of Technology, Tallinn, Estonia
2) Department of Nonlinear Geophysical Processes, Institute of Applied Physics, Nizhny Novgorod, Russia

Long linear wave transformation in the basin of varying depth is studied for a case of a convex bottom profile $h(x) \sim x^{4/3}$ in the framework of one-dimensional shallow water equation. The existence of travelling wave solutions in this geometry and the uniqueness of this wave class is established through construction of a 1:1 transformation of the general 1D wave equation to the analogous wave equation with constant coefficients. The general solution of the Cauchy problem consists of two travelling waves propagating in opposite directions. It is found that generally a zone of a weak current is formed between these two waves. Waves are reflected from the coastline so that their profile is inverted with respect to the calm water surface. Long wave runup on a beach with this profile is studied for sine pulse, KdV soliton and N-wave. Shown is that in certain cases the runup height along the convex profile is considerably larger than for beaches with a linear slope. The analysis of wave reflection from the bottom containing a shallow coastal area of constant depth and a section with the convex profile shows that a transmitted wave always has a sign-variable shape.

# 1. Introduction

The wave transformation and shoaling of the water waves in the basin of variable depth is a well developed task of fluid dynamics and has numerous applications in physical oceanography [Le Blond & Mysak, 1978; Mei, 1989; Massel, 1989; Dingemans, 1996]. Asymptotic methods are widely applied to describe the wave field for slow variations of the water depth [Shen, 1975; Mei, 1989; Dingemans, 1995; Berry, 2005; Dobrophotov et al, 2006, 2007]. In the simplified case of the 1D linear shallow-water wave propagation, asymptotic methods lead to well-known Green's law $A \sim h^{-1/4}$ for the changes of the wave amplitude A ($h$ is a water depth), derived from the energy flux conservation. Not all amplitude changes follow this law; for example, the height of a solitary wave (soliton) may vary as $A \sim h^{-1}$ in the framework of the weakly nonlinear and dispersive theory [Grimshaw, 1970; Ostrovsky & Pelinovsky, 1970]. More complicated formula can be obtained for the solitary wave of an arbitrary height (Pelinovsky, 1996). The particular law of dependence of the wave amplitude on the combination of the properties of the attacking wave and of the medium, and the related problem of wave runup is one of the central questions in tsunami modelling and modelling of flooding.

If the water depth in the coastal zone is rapidly varying, the exact analytical solutions for the wave transformation can be found within a linear shallow water theory for different bottom profiles. Such solutions usually are expressed in terms of special functions (Le Blond & Mysak, 1978; Massel, 1989; Mei, 1989). Analytical rigorous solutions of the nonlinear shallow-water system are only known to exist for only the beach of a constant slope in the vicinity of the shoreline [Carrier & Greenspan, 1958]. The solution of the nonlinear problem strongly depends on the initial wave shape. Various shapes of the periodic incident wave trains such as the sine wave (Kaistrenko et al. 1991; Madsen & Fuhrman, 2007), cnoidal wave (Synolakis 1991) and nonlinear deformed periodic waves (Didenkulova et al, 2006, 2007) have been analyzed in literature. The relevant analysis has been also performed for a variety of solitary waves and single pulses such as soliton (Pedersen and Gjevik 1983; Synolakis 1987; Kanoglu, 2004), sine pulse (Mazova et al. 1991; Didenkulova et al, 2008), Lorentz pulse (Pelinovsky and Mazova 1992), Gaussian pulse (Carrier et al. 2003; Kanoglu & Synolakis, 2006), *N*-waves (Tadepalli and Synolakis 1994) and "characterized tsunami waves" (Tinti and Tonini, 2005). Now the numerical methods are widely applied to study the wave transformation and shoaling in the coastal zone; see a review paper [Dalrymple et al, 2006].

The approximation of a linearly varying depth is not particularly realistic. Various equilibrium bottom profiles in the vicinity of the shoreline including power asymptotic $h(x) \sim x^d$ are discussed in literature. The most popular is the famous Dean's Equilibrium Profile with $d = 2/3$ (see, for example, Dean and Dalrymple, 2002). For Dutch dunes profiles with $d = 0.78$ fit better (Steetzel, 1993). Kit and Pelinovsky (1998) found a range of $d = 0.73–1.1$ for Israeli beaches. Various asymptotic approximations for beach profiles in terms of power laws are used also in theoretical models (Kabayashi, 1987; Kit and Pelinovsky, 1998). Meanwhile, in many cases, bottom profiles have composed structure and its "near-beach" general shape changes with another shape at larger depths. Fig. 1 demonstrates the bottom profile measured at Pirita Beach, Estonia (Soomere et al, 2007). It is clearly seen that the bottom profile for the depths of 2 – 10 m can be approximated by the power law with $d = 4/3$. A similar approximation with $d > 1$ can be found for continental Pacific shelf of Northern Chile for the coastal line down to the depth of 5 km (Fig. 2).

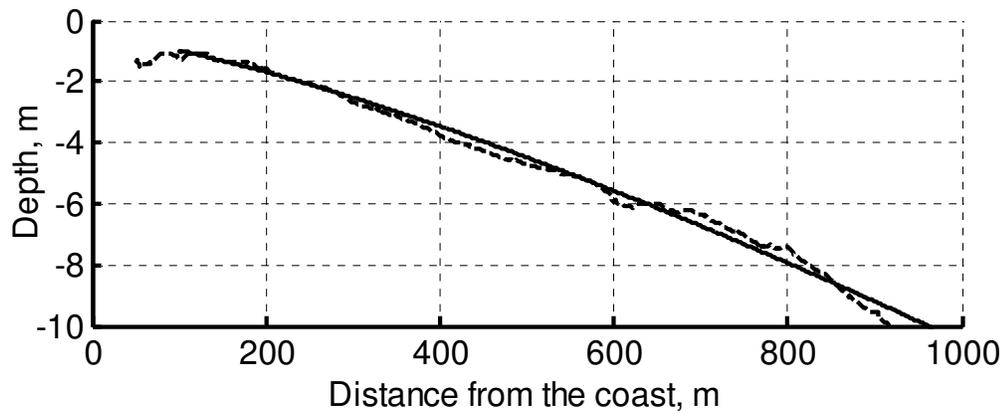

Fig. 1. Measured bottom profile at Pirita Beach, Estonia (dashed line) and its power asymptotic (solid line).

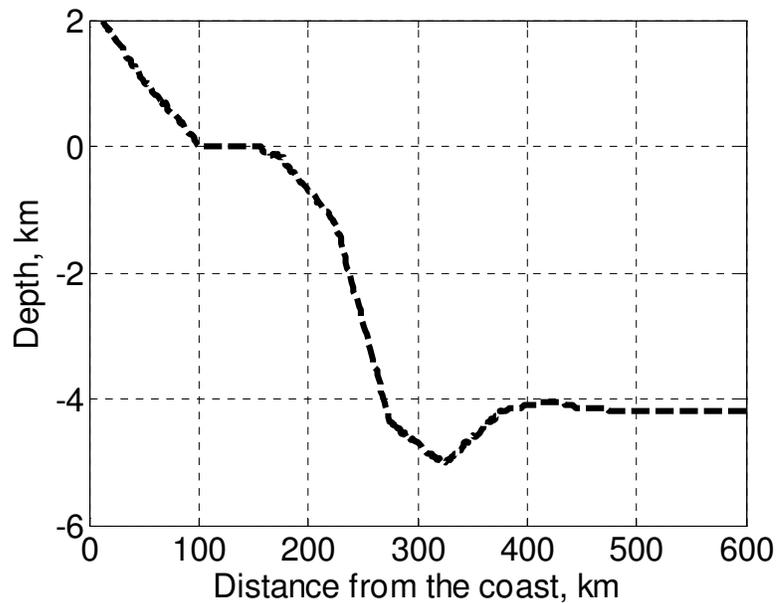

Fig. 2. Bottom profile extracted from HTDB (Gusiakov, 2002) for Pacific coast of Northern Chile (coordinates of coastal point are 7.70S 78.51W)

Therefore, the wave transformation and runup should be analyzed for various bottom profiles matching more general power laws (not only the popular case of $d = 1$). The case $d = 4/3$ has a special interest, because the solution of linearized shallow-water equations can be obtained in elementary functions for this profile [Cherkesov, 1975; Pelinovsky, 1996; Tinti et al, 2001]. In previous studies this case was considered mainly to simplify the final expressions describing wave dynamics, but a comprehensive analysis of wave properties and transformation along this type of coastal slope is missing.

In this paper we study the linear dynamics of shallow-water waves for the convex depth profile $h(x) \sim x^{4/3}$ for a wide class of initial conditions. The central goal is to establish the potential threats to the coastal zone through enhanced amplitude amplification of approaching waves and potentially larger runup height of long waves along beaches containing convex sections of the coastal slope. The paper is organized as follows. The properties of travelling waves along the convex bottom are described in section 2. The uniqueness of such travelling wave solutions is proved in section 3 by means of introducing a 1:1 transformation of the governing wave equation with varying coefficients to the constant-coefficient wave equation. This transformation enables us to obtain the solution of the Cauchy problem and to study wave evolution for various initial conditions in a straightforward manner. Wave runup on a beach of a special profile is analyzed in section 4, with an important implication that wave amplification for such a beach can be much more significant than for a plane beach. The wave propagation along the beach containing a shallow coastal area and a section of convex beach is studied in section 5. The main results are summarized in conclusion.

## 2. Travelling waves above uneven bottom

The basic model for the linear 2D shallow-water waves in the basin of a variable depth is a linear wave equation for the vertical displacement of the water surface $\eta(x,t)$

$$\frac{\partial^2 \eta}{\partial t^2} - \frac{\partial}{\partial x}\left[c^2(x)\frac{\partial \eta}{\partial x}\right] = 0, \qquad c(x) = \sqrt{gh(x)}, \tag{1}$$

where $c(x)$ is the wave speed, $h(x)$ is the water depth and $g$ is the gravity acceleration. The domain, boundary and initial conditions for Eq. (1) will be discussed later.

Travelling wave solutions for the wave equation with slowly varying coefficients, equivalently, for the waves above slowly varying bottom relief, are usually studied with the use of asymptotical methods. Solution of the wave equation for a sine wave above a convex beach with power 4/3 has the same form [Cherkesov, 1975; Pelinovsky, 1996]. We will shortly recall these results from the viewpoint of the structure of the travelling water waves.

Travelling (progressive) waves are sought in the form

$$\eta(x,t) = A(x)\exp\{i[\omega t - \Psi(x)]\}, \tag{2}$$

where $A(x)$ and $\Psi(x)$ are real functions (local amplitude and phase, respectively) which should be determined, and $\omega$ is the wave frequency. After substitution of Eq. (2) to Eq. (1) the real and imaginary parts of the resulting equation are two ordinary differential equations:

$$\left[\frac{\omega^2}{gh(x)} - k^2(x)\right]A + \left[\frac{d^2A}{dx^2} + \frac{1}{h}\frac{dh}{dx}\frac{dA}{dx}\right] = 0, \qquad (3)$$

$$2k\frac{dA}{dx} + A\frac{dk}{dx} + \frac{1}{h}\frac{dh}{dx}kA = 0. \qquad (4)$$

Here $k(x) = d\Psi/dx$ is a variable wave number. Equation (3) can be interpreted as the generalised dispersion relation for waves in inhomogeneous medium whereas Eq. (4) has the meaning of the energy flux conservation. While Eq. (4) can be easily integrated:

$$A^2(x)k(x)h(x) = \text{const}, \qquad (5)$$

Eq. (3) is a second-order differential equation with variable coefficients which generally has no analytic solutions in closed form. This equation is not simpler than the initial wave Eq. (1).

Further progress in analytical solving of Eq. (3) can be made when wave propagates above slowly varying bottom. In this case both variations of the water depth and the wave amplitude are slow; the terms in the second bracket of Eq. (3) are small compared to other additives and can be ignored in the first approximation. In this case Eq. (3) is simply solved:

$$k(x) = \frac{\omega}{\sqrt{gh(x)}}. \qquad (6)$$

Eq. (6) generalizes known dispersion relation for water waves in the basin of slowly varying depth. Solution (6) together with Eq. (5) determine the wave amplitude (that in the case of question evidently follows the Green's law) and phase. The relevant asymptotic procedure and all higher-order corrections of the wave amplitude and phase are described in detail in [Maslov, 1987, 1994; Babich & Buldyrev, 1991; Berry, 2005].

Basically, Eq. (3) can be solved numerically for an arbitrary function $h(x)$ or analytically for specific bottom profiles. As a result, solution (2) can be determined completely. Sometimes, solutions of this type are called travelling waves in an arbitrarily inhomogeneous medium (without any specific applications for water waves) and can be interpreted as a description of complicated physical processes of wave transformation and reflection in a basin of variable depth [Ginzburg, 1970; Brekhovskih, 1980].

We, however, concentrate on the analysis of the potential existence of exact travelling wave solutions to Eq. (1) and their propagation and reflection properties. The procedure to select the travelling wave solution from the entire set of solutions of Eq. (3) does not have rigorous formulation in literature. Historically, a subset of such solutions has been found by requesting that two equations

$$\frac{\omega^2}{gh(x)} - k^2(x) = 0 \tag{7}$$

and

$$\left[\frac{d^2A}{dx^2} + \frac{1}{h}\frac{dh}{dx}\frac{dA}{dx}\right] = 0. \tag{8}$$

are satisfied simultaneously. Obviously, any set of solutions $\{A, k, h\}$ to Eqs. (7) and (8) also solves Eq. (3) although generally solutions to Eq. (3) do not solve Eqs. (7) and (8) simultaneously. The solution of Eq. (7) is straightforward and given by Eq. (6); thus the function $k(x)$ is uniquely defined. The system of Eqs. (5) and (8) is overdetermined for the wave amplitude. Its consistent solution can be achieved if and only if

$$h(x) = p(x+b)^{4/3}, \tag{9}$$

where $p$ and $b$ are arbitrary constants. The desired solution therefore only exists for beaches having a specific convex bottom profile. As constant $b$ can be eliminated by a shift $\tilde{x} = x - b$ of the $x$-axis, we can assume $b = 0$ without the loss of generality. Doing so simply means that the origin $x = 0$ is located at the coastline. For the bottom profile presented by Eq. (9) the components of the travelling wave in ansatz (2) are then completely and uniquely defined:

$$k(x) = \frac{\omega}{\sqrt{gp}} x^{-2/3}, \qquad \Psi(x) = \frac{3\omega}{\sqrt{gp}} x^{1/3} + \text{const}, \qquad A(x) = \frac{\text{const}}{x^{1/3}}, \tag{10}$$

The corresponding full solution to Eq. (1) can be re-written in traveling wave form

$$\eta(x,t) = A(x)\exp\{i\omega[t - \tau(x)]\}, \qquad A(x) = A_0 \left[\frac{h_0}{h(x)}\right]^{1/4}, \qquad \tau(x) = \int_{x_0}^{x} \frac{dy}{c(y)}, \qquad (11)$$

where $A_0$ and $h_0$ are the amplitude and the water depth at the point $x = x_0$, respectively. The location of the point $x = x_0$ can be chosen arbitrarily. This feature enables analysis of waves approaching from offshore as well as waves generated in the vicinity of the coast. The solutions given by Eq. (11) correspond to right-going (propagating offshore in this geometry) monochromatic wave trains [Cherkesov, 1975; Pelinovsky, 1996]. The resulting expressions coincide with the asymptotic wave solution for the slowly varying bottom profile, but are correct for any bottom slope. A similar solution can be evidently obtained for a wave propagating to the left (onshore) direction by simply picking up another sign of $\tau(x)$ in Eq. (11). In the linear framework these waves do not interact with each other: the resulting surface displacement in areas where they excite water displacement or local current, the resulting wave profile or current speed is just sum of displacement or currents caused by the counterparts.

The previous studies into the problem in question have been limited to the analysis of properties of monochromatic or sine waves. An obvious generalisation of the existing results consists in the use of Fourier analysis to obtain the superposition of such sine waves with different frequencies, the technique obviously being applicable in this linear framework. With the use of Fourier integral of spectral components (11), travelling wave of an arbitrary shape can be presented in a general form

$$\eta(x,t) = A(x)f[t - \tau(x)], \qquad (12)$$

where $f(t)$ describes the wave shape (interpreted here as the variation with time of the surface elevation at a fixed point). An important feature is that representation (12) allows considering wave pulses of finite duration – generalized solutions of the wave equation.

Another important property of the shallow-water wave field is the wave-induced, depth-averaged flow velocity. This velocity can be calculated from the water displacement using one of equations of the linear shallow-water system

$$\frac{\partial u}{\partial t} + g\frac{\partial \eta}{\partial x} = 0. \tag{13}$$

In particular, the velocities induced by the monochromatic wave (11) and by a pulse (12) are

$$u(x,t) = U(x)\left[1 + \frac{\sqrt{gh}}{4hi\omega}\frac{dh}{dx}\right]\exp\{i\omega[t - \tau(x)]\}, \quad U(x) = A(x)\sqrt{\frac{g}{h(x)}} = A_0\sqrt{\frac{g}{h}}\left[\frac{h_0}{h(x)}\right]^{1/4}, \tag{14}$$

$$u(x,t) = U(x)\left\{f(\xi) + \frac{\sqrt{gh}}{4h}\frac{dh}{dx}\Phi(\xi)\right\}, \tag{15}$$

where $\Phi(\xi) = \int f(\xi)d\xi$ and $\xi = t - \tau(x)$. Notice that the first terms in Eqs. (14) and (15) correspond to the asymptotic solution of Eq. (1) above a slowly varying bottom, for which the shapes of the water displacement and the wave-induced water flow coincide. The second term becomes important in the vicinity of the shoreline.

The natural restriction of realistic pulses that the disturbance should have a limited energy (equivalently, finite effective wave duration) leads to the following condition

$$\int_{-\infty}^{+\infty} f(t)dt = 0, \tag{16}$$

from which it follows that the shape of a water displacement should be sign-variable. This condition as it will be shown later is valid for the travelling wave only. As it is not obvious from the viewpoint of the classical d'Alembert solution of the generic wave equation (which may consist of two sign-constant impulses propagating in different directions), we will discuss it in more detail in section 3 where the Cauchy problem will be solved.

Fig. 3 demonstrates the travelling wave propagating onshore along a coast with the bottom profile (9) with a coefficient $p = 0.01$ m$^{-1/3}$. This value will be used in all computations below. It illustrates the conservation of the shape of the water displacement and a strong deformation of the water flow. The shapes of the vertical displacement and water flow are almost identical offshore (at large depths), but they are different near the shoreline (small depth). While the wave shape remains symmetrical, the wave-induced water flow is asymmetrical: at small depths is directed offshore more than onshore (for a given shape of wave elevation). Considerable amplification of wave amplitudes occurs when such a wave approaches the shoreline. From Eqs.

(12) and (15) it follows that the amplitude of the "velocity wave" varies stronger than the amplitude of surface displacement.

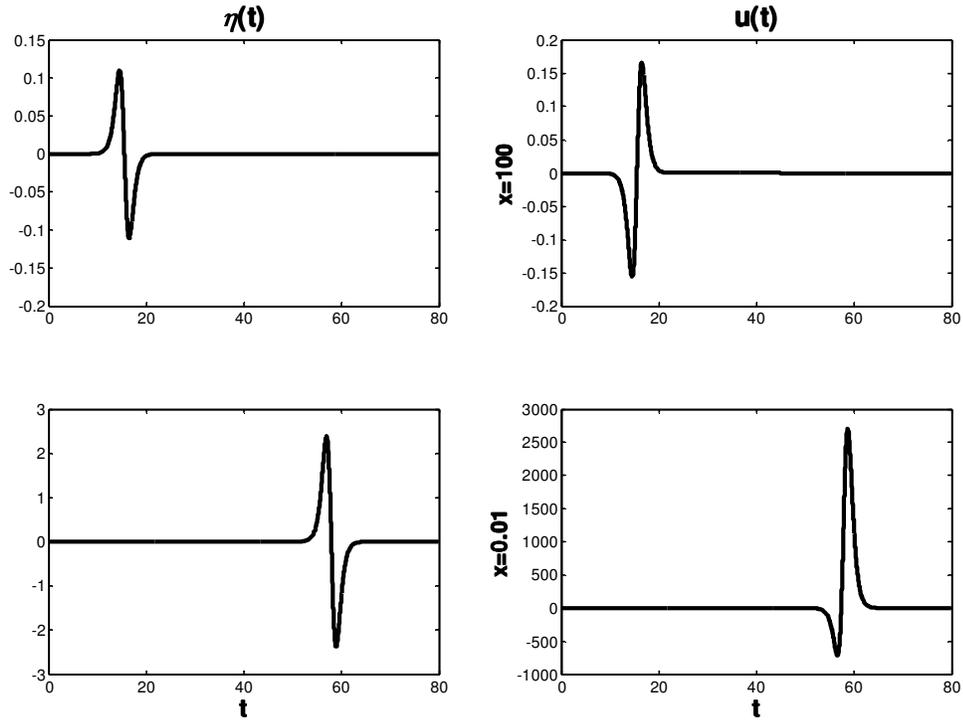

Fig. 3. The shape of a travelling wave (left) and the water flow (right) on various distances (m) from the shoreline. Time is given in sec.

## 3. Transformation to a wave equation with constant coefficients

From the form of Eq. (11) it follows that the function $f[\tau(x)\pm t]$ should satisfy a wave equation with constant coefficients. The key component of the analysis of the existence and uniqueness of solutions to Eq. (1) corresponding to travelling waves in a basin of a variable depth is establishing a 1:1 transformation of Eq. (1) to a similar equation with $c(x)=const$.

Let us seek the solution of Eq. (1) in the form

$$\eta(x,t) = B(x)H[\tau(x),t], \qquad (17)$$

where $B(x)$ and $\tau(x)$ should be determined, and the function $H$ satisfies the constant-coefficient wave equation with $c=1$, see also [Tinti et al, 2001]

$$\frac{\partial^2 H}{\partial t^2} - \frac{\partial^2 H}{\partial \tau^2} = 0. \tag{18}$$

Substitution of Eq. (17) into Eq. (1) results in Eq. (18) if and only if the unknown functions $B(x)$ and $\tau(x)$ satisfy the following three equations

$$\frac{d}{dx}\left[h(x)\frac{dB}{dx}\right] = 0, \tag{19}$$

$$h(x)\frac{dB}{dx}\frac{d\tau}{dx} + \frac{d}{dx}\left[h(x)B(x)\frac{d\tau}{dx}\right] = 0, \tag{20}$$

$$gh(x)\left(\frac{d\tau}{dx}\right)^2 = 1. \tag{21}$$

These equations are generalisations of Eqs. (5), (7) and (8). They are also overdetermined in the sense that they have a solution if and only if $h(x)$ is given by Eq. (9). In other words, the desired transformation exist if and only if the bottom profile is $h(x) \sim x^{4/3}$. This solution is unique for a reasonable choice of initial or boundary conditions and coincides with that of Eqs. (10) - (11) if $B(x) = A(x)$. Moreover, if $B(x)$ and $\tau(x)$ together with $h(x) \sim x^{4/3}$ solve Eqs. (19-21), then the transformation given by Eq. (18) reduces Eq. (1) to Eq. (18) for the unknown function $H$.

The existence of the transformation (17) if and only if the bottom profile is $h(x) \sim x^{4/3}$ proves that exact travelling wave solutions above varying bottom relief are unique to this shape of the bottom profile.

There is another important consequence from the existence of transformation (17): wave equation (18) has been extensively studied in the mathematical physics, and many theorems and approaches can be directly applied to the particular solutions in question. In what follows we use this connection for constructing the general solution of Eq. (1).

First all, wave equation (1) has a clear meaning in the given geometry and should be solved on semi-axis ($0 < \tau < \infty$) only whereas the origin $x = 0$ is a singularity point of the solution. An important simplification of the problem that the point $\tau = 0$ corresponding to the shoreline ($x = 0$) is not singular in Eq. (18). The natural boundary condition for Eq. (18) at this point is

$$H(\tau = 0, t) = 0.  \tag{22}$$

This condition implies that the water displacement $\eta(x=0,t)$ always remains bounded on the shoreline. In this case the domain for Eq. (18) can be formally extended to the whole axis ($-\infty < \tau < +\infty$). The extension is physically meaningful if the initial conditions are continued for $\tau < 0$ as $H(-\tau,0) = -H(\tau,0)$. This choice is frequently called "imaginary mirror" reflection condition. Nevertheless the wave field has a clear physical interpretation in the domain $\tau \geq 0$ only.

The general solution of the Cauchy problem for Eq, (1) describing free evolution of waves generated from the generic initial disturbance of water surface and given velocity field

$$\eta(x,0) = \eta_0(x) \qquad u(x,0) = u_0(x) \tag{23}$$

can be expressed as

$$\eta(x,t) = \frac{1}{x^{1/3}} \{ f_+[\tau(x) - t] + f_-[\tau(x) + t] - f_-[-\tau(x) + t] \}, \tag{24}$$

$$u(x,t) = \sqrt{\frac{g}{p}} \frac{1}{x} [f_+(\tau - t) - f_-(\tau + t) - f_-(-\tau + t)] - \frac{g}{3x^{4/3}} [\Phi_+(\tau - t) - \Phi_-(\tau + t) - \Phi_-(-\tau + t)], \tag{25}$$

where functions $f_+$ and $f_-$ (representing the waves propagating offshore and onshore respectively) can be found from initial conditions (23) and $\Phi_\pm(\xi) = \int f_\pm(\xi) d\xi$. The condition (22) is satisfied automatically.

In the theory of tsunami wave generation above inclined bottom only the vertical displacement in the source is usually used (Pelinovsky, 1996; Carrier et al, 2003; Tinti et al, 2005; Dutykh et al, 2006). In this case

$$f_+ = f_- = f_0[\tau(x)] = 0.5 x^{1/3} \eta_0(x), \tag{26}$$

Generally, function $f_0$ can have an arbitrary shape determined by the initial displacement. Fig. 4 displays the water displacement and velocity for the case, when the initial displacement in the source (located approximately at a depth of 20 m) is a sign-variable function (N-wave) that satisfies Eq. (16):

$$f_0(\tau) = -\frac{4s}{3}\frac{\tanh[2(\tau-60)/3]}{\cosh^2[2(\tau-60)/3]}, \tag{27}$$

where $s$ is a numerical coefficient with dimension $m^{4/3}$. In all following calculations it is taken as $s = 1$.

The initial disturbance is split into two waves after some time. The right-going wave moves quickly offshore. Its amplitude rapidly decreases and it propagates out of the domain (500 m) after 20 sec. The amplitude of the left-going wave increases as it approaches the shore. The maximum amplitude occurs at the coastline. The solution experiences perfect reflection from the shore and propagates to the right with decreasing of the amplitude afterwards.

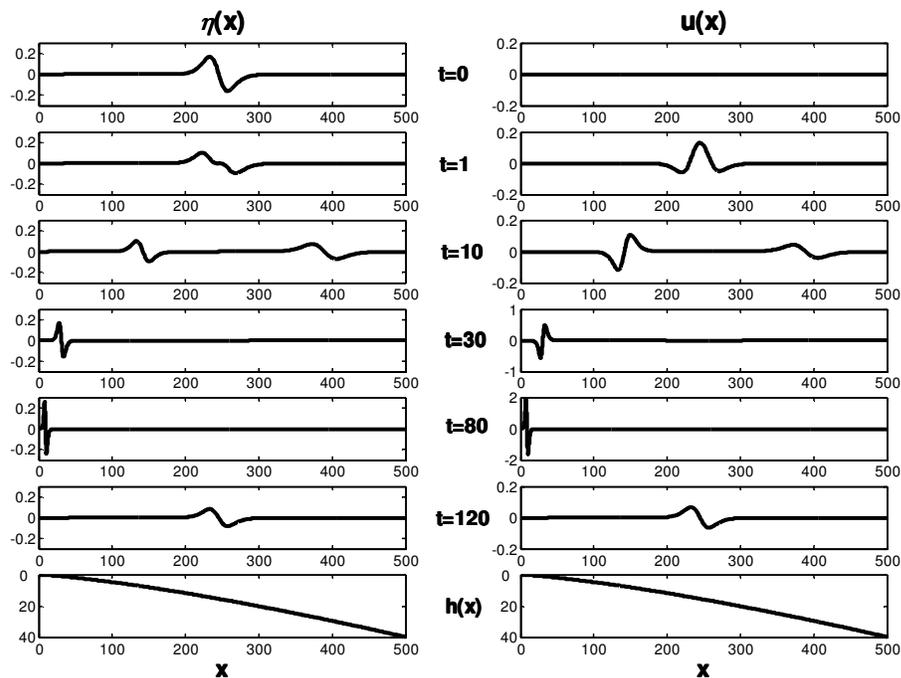

Fig. 4. Water displacement (left) and velocity (right) for initial disturbance (27). The bottom profile is shown at the bottom

Another instructive example (Fig. 5) is the propagation of initial disturbance, located entirely above the calm water level. Let us consider evolution of the wave system generated from a disturbance in the form of solitary wave

$$f_0(\tau) = s \cdot \text{sech}^2[2(\tau - 60)/3]. \tag{28}$$

An interesting feature here is the formation of a weak current between left-going and right-going pulses. It follows from the behaviour of functions $\Phi_\pm(\xi)$, not vanishing on both ends. The magnitude of this current is very small, only a few percents from the maximum velocities near wave crests (Fig. 6), thus the conservation of kinetic and potential energies is not violated.

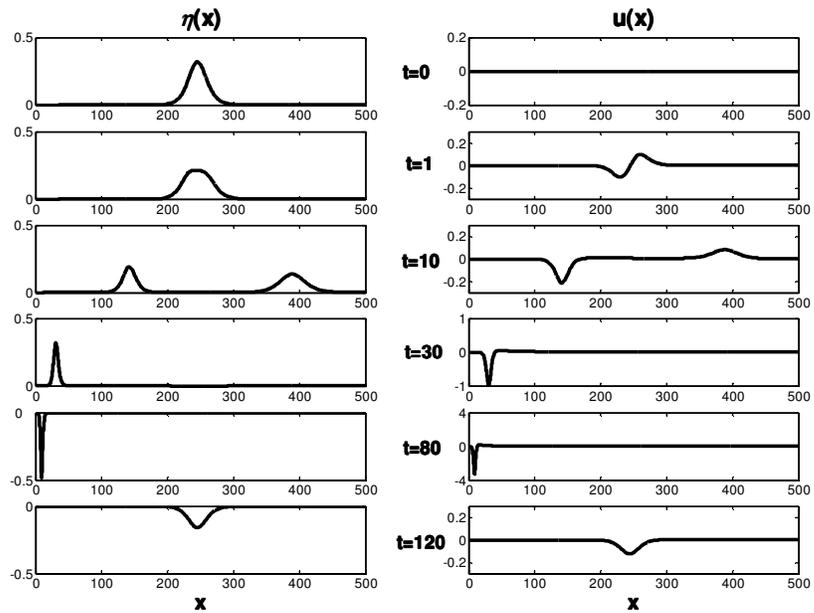

Fig. 5. Water displacement (left) and velocity (right) for initial disturbance presented by Eq. (28).

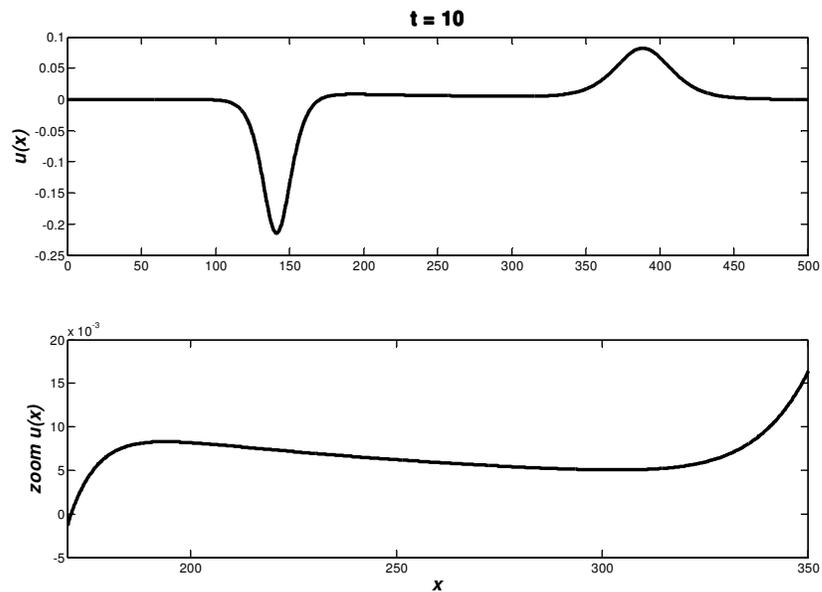

Fig. 6. Formation of space-variable current between two pulses in Fig. 5.

Initially, only positive disturbance of the water surface is present in the system. As in the previous example, the right-going wave propagates soon out of the computational domain without qualitative changes of its shape. The sign of the water elevation caused by the left-going wave, however, is inverted in the process of reflection from the coastline. After this reflection, two right-going waves exist in the system, forming together a sign-variable disturbance as expected from Eq. (11).

It is straightforward to extend the above analysis to the case of waves propagating along an ambient current. The latter can be expressed via a non-zero initial velocity field. The procedure of finding the solution is then as follows. One of the functions, for instance $f_+$, can be expressed through the initial displacement (24)

$$f_+(\tau) = x^{1/3} \eta_0(x) - f_-(\tau) \qquad (29)$$

For the other function, the following differential equation for $f_-$ (or $\Phi_-$) can be derived from (25)

$$-f_-(\tau) + \frac{1}{\tau}\Phi_-(\tau) = \Gamma(\tau) \qquad \Gamma = \frac{h^{3/4}}{2\sqrt{g}} u_0(x) - \frac{h^{1/4}}{2}\eta_0(x) + \frac{1}{2\tau}\int h^{1/4}\eta_0 dx. \qquad (30)$$

This equation can be easily integrated to give

$$\Phi_-(\tau) = -\tau \int \frac{\Gamma(\zeta)d\zeta}{\zeta}. \qquad (31)$$

The effect of the initial velocity is manifested in additional difference between the left-going (onshore) and right-going (offshore) waves.

## 4. Wave runup along a convex beach

From the practical point of view the behaviour of the wave field on the shoreline ($x = 0$) is the most interesting. Details of process of wave reflection, accompanied amplitude amplification and potential runup have important applications in tsunami modelling, forecast, and mitigation studies. Formally, the linear theory is not valid in the vicinity of the shoreline where the wave amplitude becomes comparable with the water depth. In the case of a plane beach of a constant slope it has been demonstrated that the extreme runup characteristics can be calculated rigorously from the linear shallow-water theory even for the nonlinear problem (Synolakis, 1991; Didenkulova et al., 2006, 2007). The approach used in these studies can also be applied for the case of a convex beach profile (9).

If the wave approaches the beach from the infinity, the wave solution of Eq. (1) satisfying also the boundary condition at the shoreline (22) has the following form [see Eqs. (24) and (25)]

$$\eta(x,t) = \frac{1}{x^{1/3}} \{f[t+\tau(x)] - f[t-\tau(x)]\}, \qquad (32)$$

$$u(x,t) = -\sqrt{\frac{g}{p}} \frac{1}{x} [f(t+\tau) + f(t-\tau)] + \frac{g}{3x^{4/3}} [\Phi(t+\tau) - \Phi(t-\tau)], \qquad (33)$$

where $f(t+\tau)$ is the shape of an incident wave approaching the shoreline $x=0$ ($\tau=0$). The vertical displacement of the water surface at $x=0$ can be found from Eq. (32) exactly using Taylor's series in the vicinity of $\tau=0$

$$R(t) = \eta(0,t) = \frac{6}{\sqrt{gp}} \frac{df(t-\tau_0)}{dt}, \qquad (34)$$

where $\tau_0$ is a travel time from a fixed point $x=L$ (chosen far offshore) to the shore. Taking into account that the incident wave at the point $x=L$ is

$$\eta_{in}(t) = \frac{f(t)}{L^{1/3}}, \qquad (35)$$

Eq. (34) can be re-written as

$$R(t) = 2\tau_0 \frac{d\eta_{in}(t-\tau_0)}{dt}. \qquad (36)$$

Thus, the amplitude of water level oscillations at the shoreline is proportional to the vertical velocity of water particles in the incident wave. If the incident wave has the form of a solitary crest, the water level on the shoreline experiences first runup, followed by rundown. The runup height is determined by the ratio of the travel time ($\tau_0$) to the wave period $T$. Therefore it is bigger if the incident wave approaches from deeper waters. This feature suggests that beaches that have extensive convex slopes offshore may experience considerable amplification of waves compared with beaches with linearly increasing depth.

The maximum velocity of water particles in the vicinity of the shore $x = 0$ is unbounded and proportional to

$$u(x \to 0, t) \approx 2\sqrt{\frac{g}{p}} \frac{f(t)}{x} \, . \qquad (37)$$

This feature may be interpreted as an implicit manifestation of wave breaking. However, wave breaking is not accounted for in the framework of Eq. (1). Although water velocity becomes infinitely large at the shoreline, the water discharge is bounded, because

$$h(x)u(x,t) \to 2\sqrt{gp}\, x^{1/3} f(t) \to 0 \, , \qquad (38)$$

The shore therefore plays a role of a vertical wall perfectly reflecting the wave energy from the beach.

The singularity in the water velocity in the vicinity of the shoreline can be excluded by a small variation of the bottom profile, more precisely, by variations of the face slope which is zero in a given geometry. Meanwhile, the water level is not sensitive to bottom variations in the vicinity of the shoreline. That is why we do not study detail characteristics of the velocity field at the shoreline.

To illustrate the processes in the vicinity of the coastline, time records of the water displacement during the runup of a sign-variable wave (N-wave) [Eq. (27)], computed numerically with the use of Eq. (32), are presented in Fig. 7 at selected points of the coastal slope. Far from the shoreline, the time series contain both incident and reflected waves (the latter having an inverted shape as discussed above). The wave amplification when the wave approaches the shore and the transformation of the wave shape at the shoreline are clearly seen from this Figure. For this particular wave shape, the rundown amplitude significantly (approximately in 3 times) exceeds the runup amplitude. According to Eq. (36) the maximal runup height is 5.7 m, and rundown depth is 17 m for initial amplitude 11 cm. The wave amplitude on such a beach can be amplified by an order of magnitude and even more.

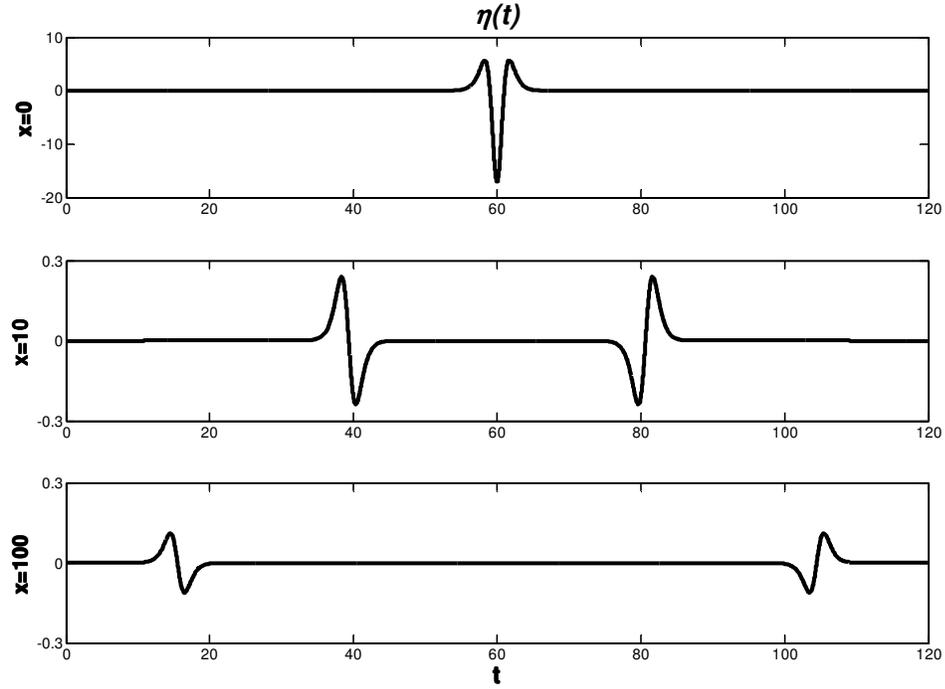

Fig. 7. Time series of the water surface of the wave system generated from initial disturbance given by Eq. (28) at the shoreline and at two offshore points

As an example, let us calculate the runup height analytically for the case when the incident wave is the solution of the Korteweg – de Vries (KdV) equation

$$\eta(t) = A\operatorname{sech}^2\left[\sqrt{\frac{3Ag}{4h^2}}t\right]. \tag{39}$$

The runup height induced by approaching solitary wave is

$$R_{max} = 4L\left(\frac{A}{h}\right)^{3/2}. \tag{40}$$

If we introduce the mean slope of a beach $\alpha = h/L$, expression (40) can be re-written as

$$R_{max} = 4\frac{A}{\alpha}\sqrt{\frac{A}{h}} \sim A^{3/2}. \tag{41}$$

Comparison of this result with the asymptotic formula for runup of a solitary wave on the plane beach (Synolakis, 1987)

$$R_{max} = 2.8312 \frac{A}{\sqrt{\alpha}} \left(\frac{A}{h}\right)^{1/4} \sim A^{5/4}. \tag{42}$$

suggests that the runup of solitary waves of moderate amplitudes on convex beaches may lead to considerably larger extent of inundation of mainland than the similar process next to beaches of a constant slope.

The runup of waves of an arbitrary shape can be studied in a similar way. Recently it has been shown in the framework of Carrier-Greenspan transformation for a plane beach that the runup height of asymmetric incident waves, which face slope exceeds back slope, is higher in comparison with the runup of symmetric waves (Didenkulova et al. 2006, 2007). This feature may be observed for beaches of various profiles and it is inherently evident from Eq. (36) for the convex beach.

## 5. Wave reflection from a zone of increasing depth

The obtained solution for a beach with a convex profile (9) contains a singularity at $x=0$. To eliminate the role of singularity, let us consider the situation when the water flow continues moving inland in a channel of small but finite depth. This situation realizes for a water channel in the port or a small river with a weak current. This can be done by considering geometry of a following bottom relief in which the origin separates a shallow area of constant depth from a convex slope (Fig. 8):

$$h(x) = \begin{cases} h_0 & x < 0 \\ h_0 (1 + x/L)^{4/3} & x > 0 \end{cases}. \tag{43}$$

In this case the velocity field is bounded everywhere. As the coastal slope is discontinuous at the origin, the presence of this inflection point gives rise to specific problem of transmission of wave energy between different areas and reflection from this point.

Let us first consider the case when an incident sine wave approaches convex coast from a zone of a constant depth ($x<0$). Following the classical theory of long wave reflection, the wave field in this zone is presented by the superposition of an incident and reflected waves

$$\eta(x,t) = A_i \exp[i\omega(t - x/c_0)] + A_r \exp[i\omega(t + x/c_0)]. \tag{44}$$

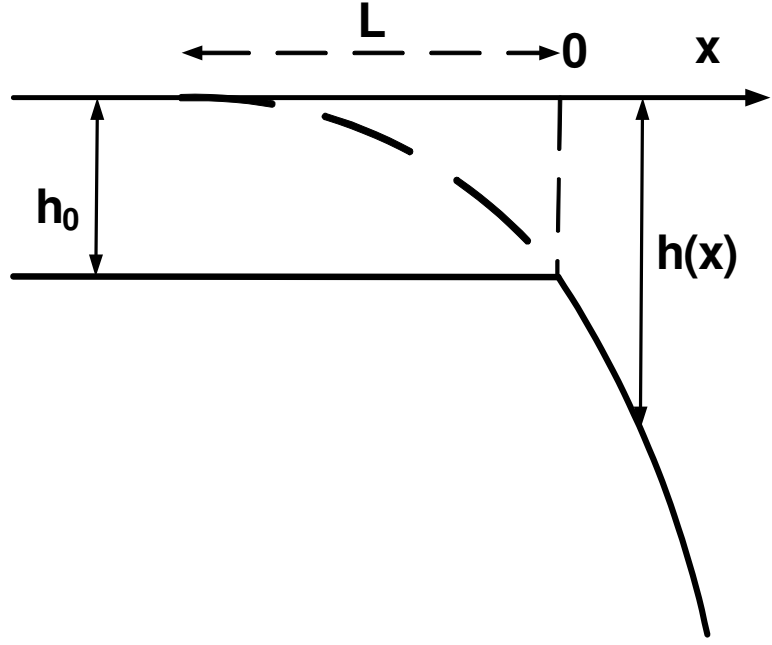

Fig. 8. Sketch of geometry of a shallow coastal area and a convex slope

Here $c_0 = \sqrt{gh_0}$ is the long wave speed on even bottom and $A_i$ and $A_r$ are amplitudes of the incident and reflected waves, respectively. The monochromatic wave along the convex slope is described by Eq. (11) and has amplitude $A_0$ at the point $x = 0$. At the inflection point the solutions expressed by Eqs. (44) and (11) must match each other in terms of the continuity of water level and total discharge. These boundary conditions allow calculating the relative amplitudes of the reflected and transmitted waves from the following expressions for the coefficients of reflection and transmission:

$$\frac{A_r}{A_i} = -\frac{1}{1 + i\omega\tau}, \quad \frac{A_0}{A_i} = \frac{i\omega\tau}{1 + i\omega\tau}, \tag{45}$$

where $\tau = 6L/c_0$. These amplitudes depend on ratio of the wave frequency (period) and the travel time of wave propagation to the zone of a variable depth. As expected, for steep bottom slopes ($\omega\tau \ll 1$) the wave is almost completely reflected and experiences a phase changes by $180^0$. For gentle slopes ($\omega\tau \gg 1$) the incident wave passes to the zone of a variable depth almost without reflection.

Another important particular case is the reflection of a solitary wave, propagating offshore. In this case Eq. (45) presents the operator form of the ordinary differential equation (that can be obtained from this equation by replacing $i\omega$ by $d/dt$):

$$\eta_r(t) + \tau \frac{d\eta_r}{dt} = -\eta_i(t). \tag{46}$$

This equation allows finding reflected wave in the vicinity of the inflection point if an incident wave in the same point is known. The details of dispersion relation transformation to differential equations in general case are described in [Whitham, 1974]). The reflected wave can be calculated as an integral

$$\eta_r(t) = -\frac{\exp(-t/\tau)}{\tau} \int_0^t \eta_i(z) \exp(z/\tau) dz, \tag{47}$$

whereas it is assumed that the reflected wave is absent before the incident wave approaches the inflection point. If the incident wave is a pulse of finite duration $T$ ($0 < t < T$) then from Eq. (47) it follows that the reflected wave amplitude in the inflection point exponentially decreases after passing the incident wave $t > T$:

$$\eta_r(t > T) = -\frac{\exp(-t/\tau)}{\tau} \int_0^T \eta_i(z) \exp(z/\tau) dz. \tag{48}$$

Sometimes it is said that the reflected waves has an exponentially decreasing tale in such cases.

From Eq. (48) it follows that the solitary wave in the water channel may entirely cross the convex slope and the inflection point without any loss of its energy. This happens for specific shapes of the incident wave and specific values of beach parameters, for which integral (48) is equal to zero. We do not include these specific cases in our analysis.

From Eq. (46) it follows that

$$\int_{-\infty}^{+\infty} \eta_r(t) dt = -\int_{-\infty}^{+\infty} \eta_i(t) dt. \tag{49}$$

Thus, if the incident wave is a wave of elevation (pure crest), a wave of depression (pure trough) dominates in the reflected wave. This feature can be interpreted as generalization of the above-discussed property of inversion of the shape in the process of reflection from the coastline.

As an example of transformation of a wave pulse with a limited duration, we consider an incident sine pulse (Fig. 9)

$$\eta_i(t) = A \begin{cases} \sin(\Omega t) & 0 < \Omega t < \pi \\ 0 & \text{out of the interval} \end{cases}. \tag{50}$$

An instructive feature of such a pulse is that it originally contains discontinuities of the surface slope that are gradually smoothed in the process of propagation. The profile of the surface elevation in the reflected wave, computed from Eq. (47), is

$$\eta_r(t) = -A \frac{q}{1+q^2} \begin{cases} 0 & \Omega t < 0 \\ \exp(-q\Omega t) + q\sin(\Omega t) - \cos(\Omega t) & 0 < \Omega t < \pi, \\ (1+\exp(q\pi))\exp(-q\Omega t) & \pi < \Omega t \end{cases} \tag{51}$$

where

$$q = \frac{1}{\Omega \tau} = \frac{\tan \theta}{8\Omega} \sqrt{\frac{g}{h}}. \tag{52}$$

In accordance with the above analysis, the reflected wave is inverted for all values of the parameter $q$ (Figs. 9 and 10). Its amplitude decreases and its tail gets gradually longer. The growth of the tail is more pronounced for gentle beaches. In the case of steep beaches the shape of the reflected wave is almost the same as for incident wave but has an opposite polarity.

Expressions (51) and (52) describe the shape of the reflected wave near the inflection point. It is straightforward to show using Fourier superposition of the spectral components [Eq. (44)], that the reflected wave preserves its shape in all distances from the inflection point.

The transmitted wave in the immediate region of the inflection point can be found from the boundary condition of continuity of water displacement

$$\eta_t(t) = \eta_i(t) + \eta_r(t). \tag{53}$$

Due to Eq. (49) condition (16) is satisfied automatically, a feature that was expected for the traveling wave solution (see section 2) and confirmed here by Eqs. (53) and (49).

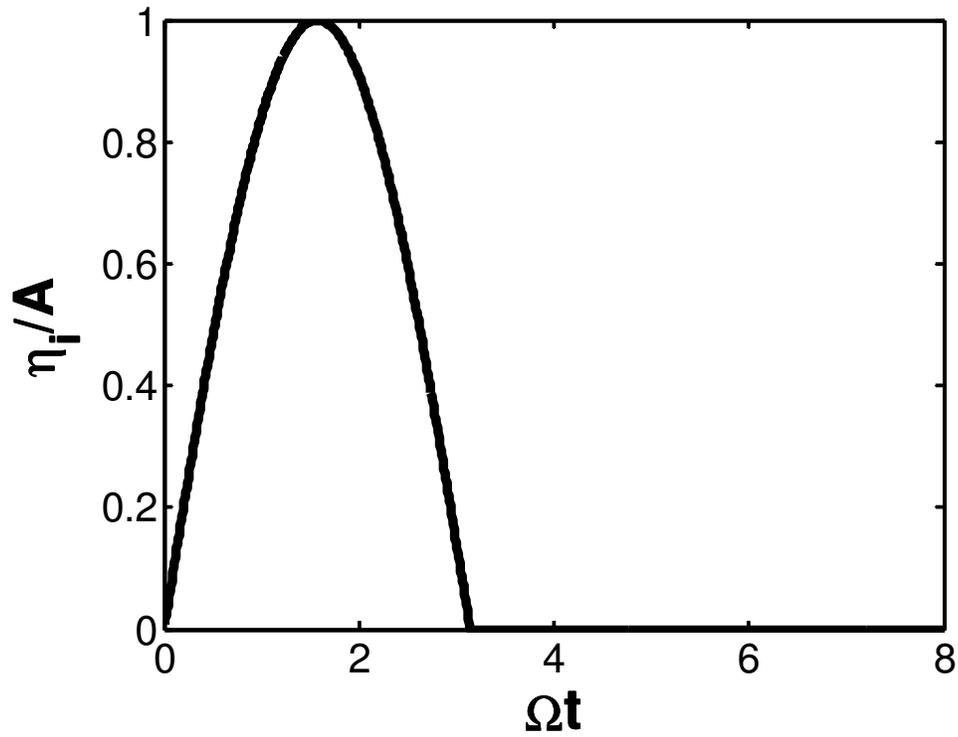

Fig. 9. The relative water surface elevation in an incident wave described by Eq. (50).

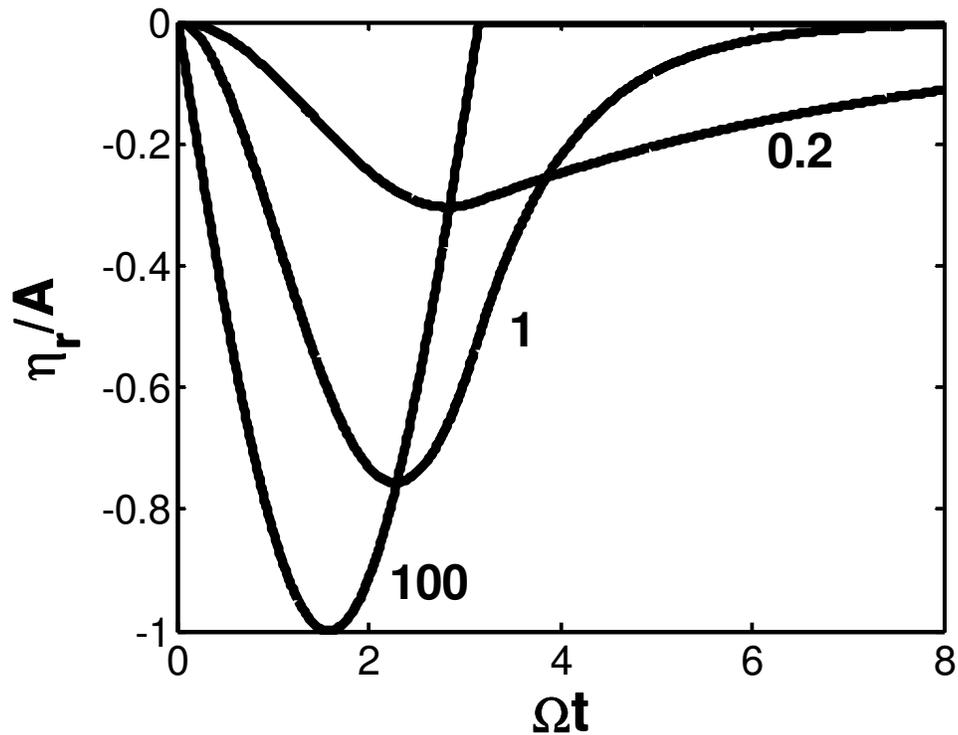

Fig. 10. The shape of reflected waves for various values of parameter $q$

The oscillations of the water level in the immediate vicinity of the inflection point are of specific influence, because they can be the starting point of studies of further description of wave attack and runup with the use of more detailed models of the coastal zone. The time series of water surface at this point are presented on Fig. 11 for the incident sine pulse. It demonstrates that a sign-variable wave is excited and propagates onshore after the inflection point. As expected, the amplitude of this wave is quite small in the case of steep convex beaches; yet almost full transmission may occur if the convex section of the beach has a moderate slope.

According to Eq. (12), the transmitted wave does not change its shape (in time), but its amplitude and phase do change with the distance from the inflection point. The shape of velocity field in a transmitted wave changes with distance as well, see Eq. (15). In the immediate vicinity of the inflection point the velocity of wave particles can be found from the boundary condition of the continuity of discharge

$$u_t(t) = \sqrt{\frac{g}{h_0}} [\eta_i(t) - \eta_r(t)]. \tag{54}$$

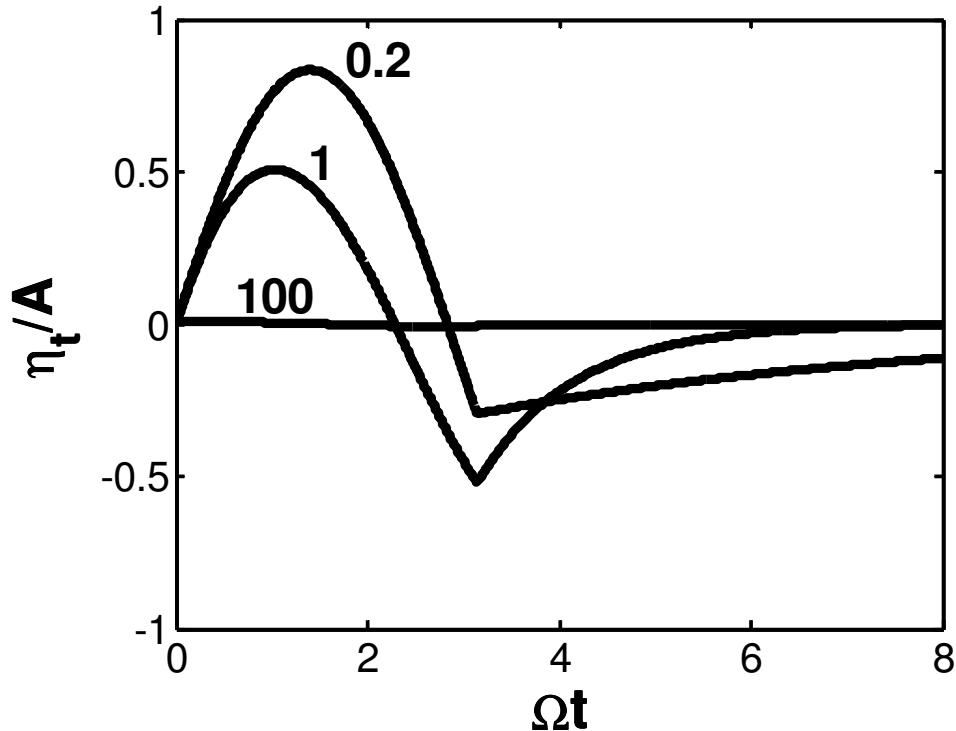

Fig. 11. The shape of transmitted wave after the inflection point for various values of parameter

$q$

Velocity time-series are presented on Fig. 12 for the case of an incident sine pulse for several values of parameter $q$. The velocity is always positive (as for incident wave). The velocity pulse is, however, somewhat modified and contains an elongated tail, the effective duration of which is larger for gentle beaches. The shape of the velocity variations in a transmitted wave varies with distance according to Eq. (15) and not necessarily follows the shape of the water surface displacements. Nevertheless far from the inflection point, the first term in Eq. (15) dominates and the shape of the velocity variations matches the shape of surface displacements. These processes are illustrated in Fig. 12.

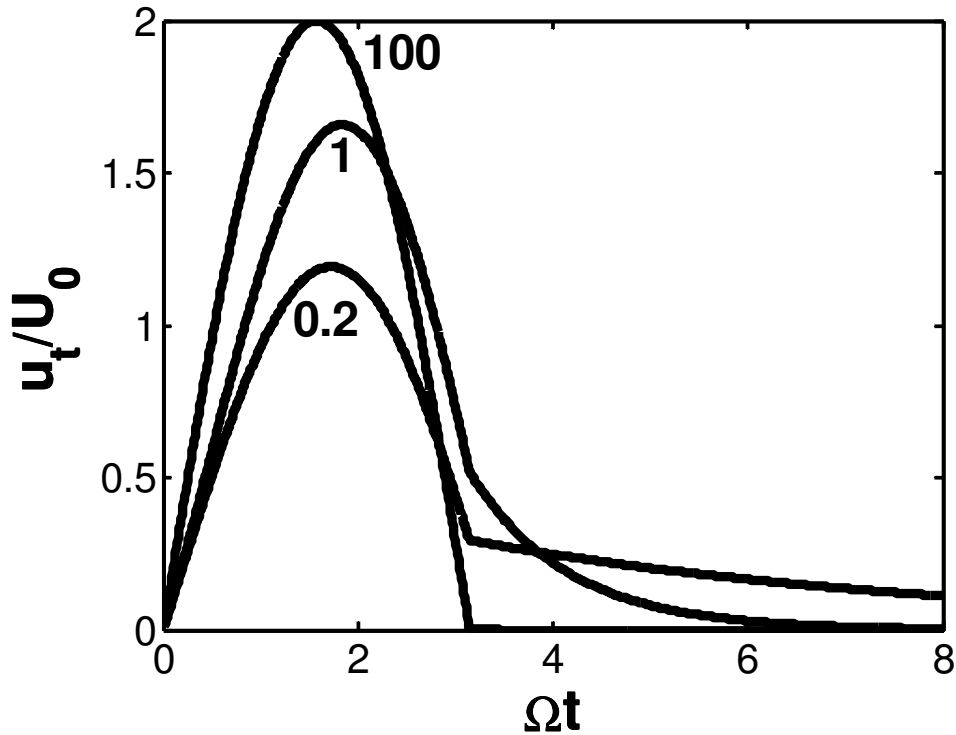

Fig. 12. The shape of transmitted velocity after the inflection point for various values of parameter $q$

## 6. Conclusion

The above analysis of linear long wave dynamics along basin of variable depth first confirms the intuitively clear opinion that travelling wave solutions of the variable-coefficient wave equation (1) exist for a very limited number of situations. In fact, such solutions exist only for a convex-shaped bottom, the water depth along which increases as $h(x) \sim x^{4/3}$. For this particular case a 1:1 transformation exists, which converts the general 1D wave equation into an analogous equation with constant coefficients. In other words, the analysed situation is the sole case in which the complex dynamics and reflections of long waves propagating over an uneven bottom can be fully described in terms of simple solutions for basin of constant depth. Notice that another generic form of the wave equation allows such solutions for another profile ($h \sim x^4$) of the coastal slope, in this case the wave equation for velocity is solved (Didenkulova et al. 2008). The obvious gain from the existence of such solutions it that quite complex wave phenomena can be analysed with the use of the large pool of results obtained for a much simpler framework. This allowed us to get important insight into how long travelling waves behave when approaching convex sections of the ocean coasts. While the majority of properties of wave propagation along a convex bottom mirrors those occurring in the basin of linearly varying depth, some interesting distinguishing features become evident; for example the shapes of water displacement and velocity in the travelling wave do not coincide. As expected, the general solution of the Cauchy problem to the wave equation in the case of the convex bottom profile in question is expressed through two travelling waves propagating in opposite directions. A zone of space-variable current generally exists between these two waves.

A deeply interesting feature is that runup of certain wave classes on a beach with this sort of bottom relief may be considerable higher than for a beach with a linear profile and with an equal mean slope. This property has been shown to hold for shallow-water KdV solitary waves (solitons) that frequently are used as a convenient model of tsunami waves. It is also shown that the shape of the water oscillations in the shoreline is determined by the first derivative of the incident wave shape. As a result, if the incident wave has a steep front, the runup height will be higher. This property is in line with recent developments in the theory of runup of asymmetric waves that indicate a strong dependence of the runup height for waves of equal height on the steepness of the face slope of the wave (Didenkulova et al. 2006; 2007).

Although initial development of the pair of waves from a wave of elevation (or depression) may lead to formation of two waves of elevation (depression), a sign-variable shape of the whole wave field is necessarily created after some time. Similarly, if the incident wave approaches a zone of increasing depth the reflected onshore-going wave has a sign-variable shape. This reflection inverts the sign of the surface disturbance and creates a wave field consisting of both elevations and depressions for any initial wave shape.

Although the exact results of the above studies are valid for a limited class of bottom profiles, they are eventually approximately correct for a much wider class of basins with a convex bottom slope, or containing extensive sections of such slopes. An important aspect to be mentioned once more is that the performed analysis does not require slow variation of the basin depth and remains valid for quite large slopes. This property allows extensive use of the obtained solutions and results for developing of practically usable models of, e.g., tsunami waves and also opens new perspectives in developing of the weakly nonlinear theory of water waves in a basin of variable depth.

**Acknowledgements**

This research is supported particularly by grants from INTAS (06-1000013-9236, 06-1000014-6046), RFBR (06-05-72011, 08-05-00069), Marie Curie network SEAMOCS (MRTN-CT-2005-019374), Estonian Science Foundation Grant 7413, and Scientific School of V.A. Zverev.

# References

**Babich, V.M., and Buldyrev, V.S.** *Short-wavelength diffraction theory. Asymptotic methods.* Springer, 1991

**Berry, M.V.** Tsunami asymptotics. *New Journal of Physics*, 2005. v. 7, 129. 18 pages

**Brekhovskih, L.M.** *Waves in layered media.* Academic Press, N.Y. 1980.

**Carrier, G.F., and Greenspan, H.P.** Water waves of finite amplitude on a sloping beach, *J. Fluid Mech.* 1958, v. 4, 97-109.

**Carrier, G.F., Wu, T.T., and Yeh, H.** Tsunami run-up and draw-down on a plane beach, *J. Fluid Mech.* 2003, v. 475, 79-99.

**Cherkesov, L.V.** *Hydrodynamics of surface and internal waves.* Naukova Dumka, 1976. (in Russian).

**Dalrymple, R.A., Grilli, S.T., and Kirby, J.T.** Tsunamis and challenges for accurate modeling. *Oceanography*, 2006, v. 19, No. 1, 142-151.

**Dean, R.G, and Dalrymple, R.A.** *Coastal processes with engineering applications*, Cambridge University Press, 2002. 475 pp.

**Didenkulova, I.I., Zahibo N, Kurkin, A.A., Levin, B. V., Pelinovsky, E.N., and Soomere, T.** Run-up of nonlinearly deformed waves on a coast, *Doklady Earth Sciences*, 2006, v. 411, 1241–1243.

**Didenkulova, I., Pelinovsky, E., Soomere, T., and Zahibo, N.** Run-up of nonlinear asymmetric waves on a plane beach. *Tsunami and Nonlinear Waves* (Ed: Anjan Kundu), 2007, 173-188.

**Didenkulova, I.I. and Pelinovsky, E.N.** Run up of long waves on a beach: the influence of the incident wave form. *Oceanology,* 2008, vol. 48, 1-6.

**Didenkulova, I.I., Pelinovsky, E.N., and Soomere, T.** 2008, Exact travelling wave solutions in strongly inhomogeneous media, Estonian J. Eng., in press.

**Dingemans, M.W.** *Water wave propagation over uneven bottom.* World Sci., Singapore,1996

**Dobrokhotov, S.Yu., Sekerzh-Zenkovich, S.A., Tirozzi, B., and Volkov, B.** Explicit asymptotics for tsunami waves in framework of the piston model. *Russian J. Earth Sciences*, 2006, v. 8, doi: 10.2205/2006ES000215.

**Dobrokhotov, S.Yu., Sinitsyn, S.O., and Tirozzi, B.** Asymptotics of localized solutions of the one-dimensional wave equation with variable velocity. 1. The Cauchy problem. *Russian Journal of Math. Physics*. 2007, v. 14, 28-56.

**Dutykh, D., Dias, F., and Kervella, Y.** Linear theory of wave generation by a moving bottom. *C.R.Acad. Sci. Paris,* Ser. 1, 2006, vol. 343, 499-504.

**Ginzburg, V.L.** *Propagation of electromagnetic waves in plasma.* Pergamon Press, N.Y. 1970.

**Grimshaw, R.** The solitary waves in water of variable depth. *J. Fluid Mech.*, 1970, v. 42, 639-656.

**Gusiakov V.M.** *Historical Tsunami Data Base,* Novosibirsk, 2002.

**Kaistrenko, V.M., Mazova, R.Kh., Pelinovsky, E.N., and Simonov, K.V.** Analytical theory for tsunami run up on a smooth slope. *Int. J. Tsunami Soc.*, 1991, v. 9, 115-127.

**Kânoğlu, U.** Nonlinear evolution and runup-drawdown of long waves over a sloping beach. *J. Fluid Mech.*, 2004, v. 513, 363-372.

**Kânoğlu, U., and Synolakis, C.** Initial value problem solution of nonlinear shallow water-wave equations. *Physical Review Letters*, 2006, v. 97, 148501.

**Kit, E., and Pelinovsky, E.** Dynamical models for cross-shore transport and equilibrium bottom profiles. *J. Waterway, Port, Coastal, and Ocean Engineering*, 1998, vol. 124, No. 3, 138 - 146.

**Kobayashi, N.** Analytical solution for dune erosion by storms. *J. Waterway, Prot, Coastal and Ocean Engineering*, 1987, vol. 113, No. 4, 401 - 418.

**Le Blond, P.H, and Mysak, L.A.** *Waves in the Ocean.* Elsevier, Amsterdam, 1978.

**Madsen, P.A., and Fuhrman, D.R.** Run-up of tsunamis and long waves in terms of surf-similarity. *Coastal Engineering*, 2007, doi: 10.1.0165coastaleng2007.09.007

**Maslov, V.P.** *Asymptotic methods of solving pseudo-differential equations.* Nauka, Moscow, 1987. (in Russian)

**Maslov, V.P.** *The complex WKB method for nonlinear equations. 1. Linear theory.* Birkhauser, Bassel, 1994.

**Massel, S.R.** *Hydrodynamics of coastal zones.* Elsevier, Amsterdam, 1989.

**Mazova, R.Kh., Osipenko, N.N., and Pelinovsky, E.N.** Solitary wave climbing a beach without breaking. *Rozprawy Hydrotechniczne*, 1991, v. 54, 71-80.

**Mei, C.C.** *Applied Dynamics of Ocean Surface Waves.* World Sci., Singapore, 1989.

**Ostrovsky, L.A., and Pelinovsky, E.N.** Wave transformation of the surface of a fluid of variable depth. *Izvestiya, Atmospheric and Oceanic Physics*, 1970, v. 6, 552 - 555.

**Pedersen, G., and Gjevik, B.** Runup of solitary waves. *J. Fluid Mech.*, 1983, v. 142, 283-299.

**Pelinovsky, E.** *Hydrodynamics of tsunami waves.* Applied Physics Institute Press, Nizhny Novgorod, 1996. (in Russian)

**Pelinovsky, E., and Mazova, R.** Exact analytical solutions of nonlinear problems of tsunami wave run-up on slopes with different profiles. *Natural Hazards*, 1992, v. 6, 227-249.

**Shen, M.C.** Ray method for surface waves on fluid of variable depth. *SIAM Review*, 1975, v. 17, 38-56.

**Soomere, T., Kask, A., Kask, J., and Nerman, R.** Transport and distribution of bottom sediments at Pirita Beach. *Estonian Journal of Earth Sciences*, 2007, v. 56, № 4, 233-254.

**Spielfogel, L.O.** Runup of single waves on a sloping beach. *J. Fluid Mech.,* 1976, v. 74, 685-694.

**Steetzel, H.J.** *Cross-shore transport during storm surges.* Delft Hydraulics, Delft, The Netherlands Publ. 1993. No. 476.

**Synolakis, C.E.** The runup of solitary waves. *J. Fluid Mech.* 1987, v. 185, 523-545.

**Synolakis, C.E.** Tsunami runup on steep slopes: How good linear theory really is? *Natural Hazards* 1991, v. 4, 221-234.

**Tadepalli, S., and Synolakis, C.E.** The runup of N-waves. *Proc. Royal Soc. London* 1994, v. A445, 99-112.

**Tinti S., Bortolucci, E., and Chiavettieri, C.** Tsunami excitation by submarine slides in shallow-water approximation. *Pure and Applied Geophysics*. 2001, v. 158, 759-797.

**Tinti, S., and Tonini, R.** Analytical evolution of tsunamis induced by near-shore earthquakes on a constant-slope ocean, *J. Fluid Mech.* 2005, v. 535, 33-64.

**Whitham, J.J.** *Linear and nonlinear waves.* Wiley, N.Y. 1974.